# On the Non-Universality of a Critical Exponent for Self-Avoiding Walks

D. Bennett-Wood†, J. L. Cardy‡, I.G. Enting¶,

A. J. Guttmann†* and A. L. Owczarek†

†Department of Mathematics and Statistics,

The University of Melbourne, Parkville, Victoria 3052, Australia

‡Department of Physics: Theoretical Physics, University of Oxford,

1 Keble Road, Oxford OX1 3NP, U.K.

¶CSIRO, Division of Atmospheric Research,

Private Bag 1, Aspendale, Victoria 3195, Australia

**Abstract**

We have extended the enumeration of self-avoiding walks on the Manhattan lattice from 28 to 53 steps and for self-avoiding polygons from 48 to 84 steps. Analysis of this data suggests that the walk generating function exponent $\gamma = 1.3385 \pm 0.003$, which is different from the corresponding exponent on the square, triangular and honeycomb lattices. This provides numerical support for an argument recently advanced by Cardy, to the effect that excluding walks with parallel nearest-neighbour steps should cause a change in the exponent $\gamma$. The lattice topology of the Manhattan lattice precludes such parallel steps.

*email: `tonyg@ms.unimelb.edu.au`

# 1 Introduction

In a recent paper, Cardy [1] using arguments based on conformal invariance theory, predicted, among other things, that the critical exponent $\gamma$ characterising the exponent of the self-avoiding walk (SAW) generating function for *oriented, interacting* SAWs is temperature dependent. (Oriented SAWs are SAWs with a direction attached to the whole walk, which in turn is associated with each step of the walk.) More specifically, if one associates a repulsive, short-ranged interaction between steps of the walk that are oriented parallel to one another, one expects to find the partition function exponent (denoted $\gamma$) changing continuously with temperature, while the radius of gyration exponent, (usually denoted $\nu$), remains unchanged. Through the hyperscaling relation $d\nu = 2 - \alpha$, we hence expect the exponent $\alpha$, characterising the self-avoiding polygon generating function, to also remain unchanged. (This is immediately obvious if one considers polygons on the square lattice, for which there cannot be any parallel neighbours).

We recently [2] presented an extensive study of oriented, interacting square lattice SAWs, based on enumerations to 29 steps. We found evidence of the predicted dependence, but unfortunately it was not a dramatic effect, in that the change $\gamma(T = 0) - \gamma(T = \infty)$ was about 0.01, or less than 1%. This is the change in the critical exponent if we exclude all walks with any parallel interactions.

It follows that if there exists a lattice on which neighbouring parallel steps are not possible we should encounter the same order of magnitude change in the critical exponent, since the lattice topology now introduces the same effective repulsion as in infinitely repulsive square lattice walks. Two two-dimensional lattices which have this property are the Manhattan lattice and the L-lattice. (In Appendix B we show how the arguments of Ref. [1] may be generalised to these lattices). For the Manhattan lattice, the SAW series were previously known to 28 terms [3], and for the L-lattice [4] to 44 terms. An early analysis [5] gave $\gamma(\text{Manhattan}) \approx 1.320$ and $\gamma(\text{L}) \approx 1.350$. These estimates are obviously too imprecise to shed any light on the problem.

On the other hand, several groups of authors [6, 7, 8, 9, 10, 11] have subsequently studied the problem of interacting oriented SAWs. They have extended the original exact enumeration study [2] with careful transfer matrix and Monte Carlo studies. While clarifying the phase diagram they find little evidence for the prediction of Cardy [1] concerning the continuously changing exponent $\gamma$. In addition they attempted to provide some theoretical explanation for this lack of confirmation of the theory. Below we discuss these arguments and provide an alternative reason for the numerical results. We also examine the exact enumeration and series analysis for the related problem (as mentioned above) of SAWs on the Manhattan lattice in another attempt to provide numerical evidence of a shift in the exponent $\gamma$. The results provide good support for Cardy's theory, though the evidence is not incontrovertible.

Accordingly, we have substantially increased the length of the series for both SAWs and polygons on the Manhattan lattice. For the SAWs, we have used a straightforward back-tracking algorithm, but implemented this on a small Intel Paragon supercomputer, enabling enumerations to 53 steps to be calculated in a reasonable time. For the polygons, we have used the finite-lattice method, as described in [12], in which 48 step polygons were obtained, but improved as described below. The improved algorithm, plus advances in computers in the intervening decade, allow polygons of perimeter 80 steps to be enumerated. By showing that the first incorrect term given by the finite lattice method is incorrect by an amount given by the number of convex polygons on the square lattice, which are exactly known, we have obtained an additional term, so that we have 84 step enumerations.

The primary purpose of the polygon enumeration is that the long series enables a quite precise estimate of the connective constant to be made. This is then used in the analysis of the SAWs. The biased analysis of the SAWs then permits a more precise estimate of the exponent $\gamma$ to be made. In this way we find

$$\mu = 1.733535 \pm 0.000002 \quad \text{and} \quad \gamma = 1.3385 \pm 0.003. \tag{1}$$

As the critical exponent for square lattice SAWs [13] is $43/32 = 1.34375$ it seems likely that the critical exponent for the Manhattan lattice is different, supporting and making more precise the observations of [2]. A more comprehensive analysis, based on the asymptotic form of the susceptibility coefficients, strengthens this conclusion.

The make-up of the paper is as follows. In the next section we describe the enumeration of SAWs, and in the following section the enumeration of polygons. The subsequent section gives an analysis of the series. In section 5 we reanalyse extended square lattice oriented walk data with the same methods used in section 4 for the Manhattan lattice data. The final section discusses the current state of the theory in relation to the new results we provide. In the appendix we prove that the correction term to the polygon enumerations is given by square lattice convex polygons.

## 2 Enumeration of Self-Avoiding Walks

The enumeration of the SAWs was carried out on an Intel Paragon supercomputer. The Paragon consists of 55 independent compute nodes (processors) which each have their own memory and are connected via a high-speed 'node interconnect network'. This means that for most applications, you can think of each node as being fully connected to all other nodes.

By exploiting the symmetry of the Manhattan lattice, we were able to divide our enumerations up among the available processors by programming each of the 139 distinct configurations of length 9 into each processor. (As we have only 55 processors available, three separate runs were make using 46, 46 and 47 nodes respectively). Each of these runs produced a speed-up of about 38, resulting in a parallelisation of nearly 81%. (If the Paragon had the full 139 nodes available, a speed-up of 114 would have been possible).

Each processor counted the number of walks $w_n$ for $n = 10 \ldots 53$ whose configurations contained one of the 139 nine-step patterns at the beginning of the walk using a simple backtracking algorithm [4]. Finally, global additions between processor resulted in the final totals. Because of the feature of very little communication throughout the algorithm, our algorithm is nearly fully parallelised.

In this way we obtained the enumerations shown in Table 1.

## 3 Enumeration of Self-Avoiding Polygons

The Manhattan lattice polygons are enumerated using the finite lattice method as described by Enting and Guttmann [12]. The program has been subjected to some minor technical modifications, drawing on improvements made in the course of our more recent studies of square lattice polygons. These are: (i) use of 'base-3' indexing for representing bond configurations, to avoid overflow problems at large widths; (ii) data-modularisation to aid transfer of the program between machines with different hardware limitations; (iii) restart capability; and (iv) dynamic allocation of storage in the transfer matrix calculations, so that storage from the 'old' vector is released as early as possible and made available for building up the 'new' vector. However the basic algorithm remains as described by Enting and Guttmann [12], i.e. a direct generalisation of the original polygon enumeration procedure [14], taking account of the different Manhattan lattice vertex types and their respective restrictions on possible bonds.

The various changes to the programs are directed at saving memory and most likely involve small penalties in run-time. The improvements in run time relative to the 1985 study reflect improved computer hardware. Of the space-saving modifications, the dynamic allocation of storage is of particular importance for the Manhattan lattice enumerations. The constraints on bond directions mean that many of the configurations needed on the square lattice can not occur. However as the 'transfer-matrix' technique adds one site at a time, the partial lattice includes one vertical bond (a different one at each successive step) and so the requisite sub-set of square lattice configurations differs at each step. In this circumstance, dynamic allocation of storage provides a significant saving.

As in previous polygon enumerations, the finite lattice method determines polygons that can be embedded in finite rectangles of width $W$ bonds and length $L$ bonds. (Note that only odd values of $W$ and $L$ are relevant for the Manhattan lattice). An inclusion-exclusion relation (with weights defined in [12]) is used to combine the contributions of finite rectangles to give the polygon enumeration for the unbounded system. The length of the series is determined by the maximum achievable value of $W$, $W_{\rm max}$. If we use $L \leq 2W_{\rm max} + 2 - W$ then the polygons are enumerated up to $4W_{\rm max} + 4$. The count for polygons of perimeter $4W + 8$ is incorrect by an amount exactly equal to the number of *convex* polygons on the *square* lattice of perimeter $2W + 6$. This is proved in the appendix.

For $W_{\rm max} = 19$, which gives 80 step polygons, the program took 12 hours on a 300 Mhz DEC Alpha. The program needed to be run four times, as we performed all calculations *modulo* a prime number less than $2^{15}$, in order to save memory. After four runs, with four different primes, integers up to $2^{60}$ can be uniquely reconstructed using the Chinese remainder theorem. In this way we obtained the counts $p_n$ shown in Table 2. Note that only polygons whose perimeter is an integral multiple of 4 may be embedded on the Manhattan lattice.

## 4 Series Analysis

We have analysed the SAW series by the method of differential approximants, evaluating both unbiased and biased approximants. The method used is precisely that described in [15]. From unbiased approximants we estimate $\gamma = 1.332 \pm 0.005$ and $1/\mu = 0.57684 \pm 0.00001$. This is a combination of results obtained from both first and second order differential approximants. As usual, the quoted errors represent twice the standard deviation of the estimates of the quantities estimated, and as such are not rigorous bounds. It is significant that the estimate of $\gamma$ is about 1% lower than the corresponding estimate for square lattice SAWs, obtained from a series of the same effective length.

A similar unbiased analysis of the polygon series resulted in the estimates $1/\mu = 0.576856 \pm 0.000003$ and $\alpha = 0.4996 \pm 0.0026$. If we bias the approximants by imposing the value $\alpha = 0.5$ and impose the linearity that is observed to exist between estimates of the singularities and exponents, we then estimate $1/\mu = 0.5768560 \pm 0.000001$, which we take to be our final estimate.

Using this estimate of $\mu$ to bias estimates of $\gamma$ we obtain $\gamma = 1.337 \pm 0.002$ from first-order approximants, $\gamma = 1.339 \pm 0.003$ from second-order approximants, and $\gamma = 1.3385 \pm 0.003$ from third-order approximants. These results, taken at face value, would preclude the conclusion that $\gamma = 43/32 = 1.34375$. However there is a trend in the data that weakens this conclusion. If we had only 34 terms of the series, our estimate of $\gamma$ from second- and third-order approximants would be around 1.336. With 44 terms this would increase to 1.3375 or 1.338 and with 54 terms our estimate, as noted above, is 1.339. It is clearly possible that with an arbitrarily long series this trend could push up the estimate of $\gamma$ to the value known to hold for the square lattice, viz. 1.34375.

For the square lattice, for which the longest series are available, a similar trend is observable. Using the 51 term square lattice series given by [16], we find that the biased estimate of $\gamma$ is 1.34365 with a 31-term series, 1.34370 with a 41 term series and 1.34375 with a 51 term series. We also see that the many higher order approximants are defective, but that that doesn't seem to affect the accuracy of the estimates. A similar pattern is observed for the Manhattan walk series.

The most likely explanation of this phenomenon is that we are fitting very long series to a functional form that *we know* is incorrect. That is to say, we have clearly demonstrated that the generating function for SAWs is not differentiably finite [16], yet we are forcing the coefficients to fit a differential equation of that form. With fewer coefficients this is numerically easier to do, but with very long series the method of analysis is essentially telling us that this is the wrong underlying form. These considerations apply also to the polygon generating function, but less forcefully, due to its simpler algebraic structure. Thus while we can actually show that the generating function for polygons is not differentiably finite, its singularity structure is sufficiently simple — there being no non-analytic correction-to-scaling exponent, that the D-finite approximants can better represent the underlying function. While the above remarks are clearly speculative, they are both consistent with our observations and the only plausible explanation we can put forward for the observed behaviour.

In [16] we used an alternative method of analysis which relied on assuming the underlying asymptotic form, which is a weaker assumption in some sense than assuming that the underlying solution is D-finite. Accordingly, we repeat that analysis *mutatis mutandis* for the Manhattan SAW generating function.

Since the polygon generating function has non-zero coefficients for polygons of perimeter $4n$, it follows that the SAW generating function, which has non-zero coefficients for SAWs of all lengths, has four singularities on the circle $|x| = x_c$, at $x = \pm x_c, \pm \imath x_c$.

It follows that the generating function behaves like

$$\begin{aligned} C(x) &= \sum c_n x^n \sim A(x)(1 - x/x_c)^{-\gamma}[1 + B(x)(1 - x/x_c)^{\Delta} + \cdots] \\ &+ D(x)(1 + x/x_c)^{\frac{1}{2}}[1 + E(x)(1 + x/x_c)^{\Lambda} + \cdots] \\ &+ F(x)(1 + x^2/x_c^2)^{\Theta}[1 + \cdots] \end{aligned}$$

where $A, B, C, D, E, F$ are smooth functions.

In our analysis of square lattice SAWs, we found the sub-dominant exponents $\Delta = \frac{3}{2}$ and $\Lambda = 1$. As we have no reason to believe that these are non-universal, and even if they were, they are unlikely to be too different from the square lattice SAW values (given that the apparent difference in the leading exponent is less than 1%) we will assume these values for the Manhattan SAW series too. The last term, which is not present in the case of square lattice SAWs follows from the fact that polygons only occur with perimeter $4n$. Careful study of the unbiased approximants to the walk generating function indicates the presence of this conjugate pair of singularities, with exponent $\Theta \approx 0.5$. A biased analysis, assuming that the critical point is at $\pm \imath x_c$ gives $\Theta = 0.54 \pm 0.04$.

Hence the asymptotic form of the coefficients is given by:

$$\begin{aligned} c_n x_c^n &\sim n^{\gamma-1}[a_1 + a_2 n^{-1} + a_3 n^{-\Delta} + a_4 n^{-2}] \\ &+ (-1)^n n^{-\frac{3}{2}}[b_1 + b_2 n^{-\Lambda} + b_3 n^{-1} + b_4 n^{-2}] \\ &+ (-1)^{\frac{n}{2}} n^{-\Theta-1}[f_1 + f_2 n^{-1} + f_3 n^{-2}]. \end{aligned}$$

for $n$ even.

In the absence of any reason to suspect the contrary, we assume only analytic corrections to the contribution of the singularities on the imaginary axis. For $n$ odd, a similar expansion holds with $\frac{n}{2}$ replaced by $\frac{n+1}{2}$. We will consider the odd- and even-subsequence of SAW coefficients separately. For $n$ even we have

$$c_{2n}x_c^{2n} \sim n^{\gamma-1}[a_1^e + a_2^e n^{-1} + a_3^e n^{-\Delta} + a_4^e n^{-2}] + (-1)^n n^{-\Theta-1}[f_1^e + f_2^e n^{-1} + f_3^e n^{-2}].$$

For $n$ odd we have

$$c_{2n+1}x_c^{2n+1} \sim n^{\gamma-1}[a_1^o + a_2^o n^{-1} + a_3^o n^{-\Delta} + a_4^o n^{-2}] + (-1)^n n^{-\Theta-1}[f_1^o + f_2^o n^{-1} + f_3^o n^{-2}].$$

Note that the antiferromagnetic singularity has "folded in" to the ferromagnetic singularity, as $\Delta = 1.5$. So the amplitudes for the odd- and even subsequence will not all be the same. To be more precise, they will differ by a factor of $x_c$ by their definition, and after taking this fact into account, $a_3^e$ from the even subsequence should not be equal to $a_3^o$. They should in fact differ by an amount equal to $2b_1$. Similar comments apply to the higher order amplitudes. However our purpose here is to identify $\gamma$, the leading amplitude, so we will not dwell on these less important aspects, except to point them out.

A fit to the even subsequence of the SAW generating function, with $1/\mu = 0.5768563$ — which is just about at the centre of our estimated range — and with $\gamma = 1.3385$, $\Theta = 1.54$, $\Delta = \frac{3}{2}$ and $\Lambda = 1$ as explained above, is shown in Table 3.

Convergence is seen to be very satisfactory. Note in particular the four digit stability of the estimates of $a_1^e$, the two digit stability of the estimates of $a_2^e$, while estimates of $a_3^e$ and $a_4^e$ are slowly oscillating around central values of $-1.6$ and $1.2$ respectively. The estimates of $f_1^e$ are stable to three digits, while $f_2^e, f_3^e$ are fairly constant.

In Table 4 we give the corresponding results for the odd subsequence.

Convergence is again seen to be satisfactory. As for the even subsequence we observe four digit stability of the estimates of $a_1^o$, the two digit stability of the estimates of $a_2^o$, while estimates of $a_3^o$ and $a_4^o$ are slowly oscillating around central values of $-2.6$ and $2.1$ respectively. The estimates of $f_1^o$ are just about stable to three digits, while $f_2^o$ is fairly stable and estimates of $f_3^o$ are slowly oscillating.

In Table 5 we give the corresponding results for the even subsequence, with the sole change that the estimate of $\gamma$ is set to $\frac{43}{32}$, the square lattice SAW value.

Convergence is seen to be substantially worse than that seen in the tables above. We observe a smooth downtrend of the estimates of $a_1^e$, a fairly strong uptrend of the estimates of $a_2^e$, while estimates of $a_3^e$ and $a_4^e$ are clearly diverging. The estimates of $f_1^e$ display a two period oscillation, *which indicates that the ferromagnetic singularity is not quite correct.* This is reinforced by the more strongly oscillatory behaviour of $f_2^e$ and $f_3^e$.

Thus with these values of the parameters, it is clear that $\gamma = 1.3385$ is strongly preferred over $\gamma = 1.34375$. We now show that this conclusion does not change if we alter the various parameters within a reasonable range.

In Table 6 we give the results for the even subsequence, with the sole change from Table 3 being that the estimate of $x_c = 3.005133$.

It is apparent that while the numerical values of the amplitudes change slightly, the overall quality of the fit is largely unchanged. The poorer fit observed with a change of $\gamma$ to the square SAW value persists with a change of critical point, (though to save space we don't tabulate that fit.)

Our final table shows the result of changing our exponent estimate of $\Theta$ from the observed 1.54 to the more appealing simple rational fraction $\frac{3}{2}$.

The stability of the amplitude estimates is seen to be slightly worse than the corresponding results in Table 3, and in particular we see a slightly oscillatory trend in the amplitudes $a_2^e, a_3^e, a_4^e$ which is characteristic of an incorrect value for $\Theta$.

A variety of other combinations of values for the various parameters were tried, but these did not alter our conclusions based on the above tables.

We conclude that we have persuasive numerical evidence in favour of the critical exponent $\gamma$ for Manhattan lattice SAWs being different to the corresponding result for square lattice SAWs. While the difference is small — just less than half of 1%, it is nevertheless seemingly present. It would be most valuable to have a further 10-20 terms of the Manhattan SAW series, but even 10 further terms would require an increase in computer time by a factor of 250 using the present algorithm.

## 5 Re-analysis of Oriented Square Lattice Data

We have taken this opportunity to reanalyse the data for oriented square lattice walks at $\beta = -\infty$, partly to allow a comparison with the Manhattan data analysis, and partly to make use of the 5 extra series coefficients that were recently obtained by Barkema and Flesia [8]. We note in passing that the first new coefficient for 30-step walks with no parallel contacts given by Barkema and Flesia [8] is wrong. It was obtained by subtracting the number of 30-step walks with at least 1 parallel contact from the total number of walks (given in [17]), and it appears that this subtraction was erroneously reported. The correct coefficient should be, for 30-step walks with no parallel contacts, 4113237913603.

With this correction, we reanalyse the data using the best estimate for the critical point, as previously reported in Bennett-Wood *et. al.* [2] and the exact exponents for unoriented SAWs, as given by Conway and Guttmann [16]. The results are shown in Table 8 below. Note that there is no four-term periodicity, coming from the formation of polygons, in the square lattice SAW. Hence, the full sequence, and not just odd or even sub-sequences, have been utilised in the analyses, and also displayed in all the tables of this section.

It can be seen that estimates of the leading amplitude are dropping in the 4th decimal place, and both the sub-leading amplitude estimates are changing in the second decimal place. More significantly, the antiferromagnetic amplitudes are clearly oscillating, which indicates that the ferromagnetic singularity is not correct. The oscillations in the sub-leading antiferromagnetic exponent are particularly severe.

In Table 9 we give the corresponding results with $\gamma = 1.3394$. Given that the antiferromagnetic exponent should be related to the exponent $\alpha$, which is not expected to vary with the strength of the parallel interactions, we use the value 0.5 for this exponent.

The improvement in the fitted estimates is considerable. The leading amplitude is changing in the 5th decimal place, the subleading amplitude in the 4th decimal place, and the confluent amplitude is changing in the 3rd decimal place. More significantly still, the antiferromagnetic amplitudes are also quite stable, and show no indication of an oscillation, which would be the hallmark of an incorrect ferromagnetic exponent.

Finally, we repeat the above analysis, *mutatis mutandis* for oriented two-legged stars on the square lattice. Unfortunately we only have a comparatively short series of 27 terms. For unoriented two-legged stars, the ferromagnetic and antiferromagnetic exponents are increased by exactly 1 over their one-legged star (SAW) counterparts. As we have discussed previously [2], we expect the change in exponent at $\beta = -\infty$ for two-legged stars to be three times as large as the corresponding change for SAW.

In Table 10 we give our results for the value of the exponent that leads to the most stable estimates for the amplitudes. This turns out to be $\gamma^{(2)} = 2.3215$. This change is in fact nearly 5 times the corresponding change in $\gamma$. We do not regard this as significant. Rather the contrary. The fact that this analysis is sensitive enough to point out that the exponent shift in two-legged stars is substantially greater than that for SAW is a significant endorsement of both the underlying theory and our method of analysis.

Thus in summary we find, for oriented square lattice SAW with no parallel contacts an exponent of $1.3395 \pm 0.002$ and for oriented two-legged stars with no parallel contacts an exponent of $1.3215 \pm 0.005$. Thus the shift from the unoriented counterparts is $0.004 \pm 0.002$ — which is comparable to that found for Manhattan SAW of $0.005 \pm 0.003$ — and for two-legged stars of $0.022 \pm 0.005$.

## 6 Discussion

It is now the case that analysis of exact enumeration on both the square and Manhattan lattices has estimated the change in the exponent $\gamma$ to be of the order of 0.005 in both cases. In contrast, Monte Carlo and some transfer matrix studies [6, 8, 9, 10, 11, 18] seem to be consistent with no change at all. In particular, the predicted variation of $\gamma$ with the parallel interaction on the square lattice implies that, in the absence of such an interaction, the mean number of parallel contacts should grow as $\ln N$, with a coefficient of the order of $\gamma'(0)$. Analysis of the Monte Carlo data has led the authors of Refs. [6, 8, 10, 11] to the conclusion that this behaviour is not exhibited: rather the mean number of parallel contacts appears to saturate.

It may at first sight seem surprising that methods based on series expansions, which are exact only for the first few terms, can claim to give results as good as, or better than, Monte Carlo analysis that may involve walks of $10^5$ or even more steps. However if the correct asymptotic form is accounted for, that is indeed the case. As an example we cite the situation that prevails for ordinary SAW on the square lattice. In [16] a 51 term series for the walk generating function is derived and analysed to give, for the critcal exponent, $\gamma = 1.34367 \pm 0.00010$ and $\gamma = 1.34372 \pm 0.00010$ from two different methods of analysis - that is, an apparent error of better than 0.01% in both cases. By way of comparison, a state of the art Monte Carlo analysis, based on walks of up to 80000 steps is reported in [19]. While they didn't estimate $\gamma$ itself, they did estimate $2\Delta_4 - \gamma$, for which they found the value $1.4999 \pm 0.0002$, an error of a little more than

0.01%.

Our purpose here is not to advance the claims of one method over the other, but rather to point out that they are complementary. When they agree, their (individual) conclusions are strengthened. When, as in the present case, they disagree, it is appropriate to suspend judgement unless one method can be shown to be inappropriate. For the present problem, both methods have weaknesses. The key to this predicted effect can be restated as a question. Is it the case that the mean number of parallel contacts, in the case of the square lattice, grows with $\ln N$? From series work on necessarily short walks it appears that it does. Longer walks, studied by Monte Carlo methods, suggest that the linear behaviour eventually saturates, and that the mean number of contacts grows much more slowly. The field-theoretic argument given toward the end of this section provides one explanation for this phenomenon. Further, we argue that the true asymptotic regime, where this effect will be clear, requires walks of length rather greater than $10^9$ steps — far outside any possible Monte Carlo or series work. Thus we could surmise that the series approach has captured the correct behaviour as it picks up the first occurrence of contacts, and the Monte Carlo method has got it wrong as it picks up merely repeats of these contacts, rather than independent new contacts. Or we could argue that the series are clearly far too short to be believed, and that so are the Monte Carlo simulations, though they are longer than the series! Such speculations are, ultimately, profitless. Rather, we prefer to let our data and analysis speak for itself, and simply claim that we have found evidence in favour of the predicted effect, but not overwhelmingly persuasive evidence.

There have been no detailed refutations of the original arguments [1] that led to the conclusion that $\gamma$ should indeed change as the parallel contact energy is varied. An argument advanced in Ref. [10], based on an interpretation of scale invariance, is, in our opinion, flawed. These authors point out that scale invariance of long SAWs in the plane labelled by polar coordinates $(r, \theta)$ implies that they are translationally invariant in $\ln r$. They then assert this means that there are the same mean number of parallel "close encounters" in each annulus $\ln r_0 < \ln r < \ln r_0 + \Delta$, implying that there is a uniform density of such close encounters along the $\ln r$ axis. However, these authors assert, parallel contacts on neighbouring lattice bonds correspond to $\Delta \sim 1/r$, so that the total number of such contacts is of the order $\int_0^{\nu \ln N} (1/r) d(\ln r)$ which is finite as $N \to \infty$. This argument, while appealing, is flawed because (a) the notion of scale invariance only applies to properties defined on distance scales much larger than, and independent of, the lattice spacing, since the latter introduces a length scale which clearly violates scale invariance; (b) the term "parallel close encounters" is ill-defined, while the notion of parallel contacts clearly refers to a lattice, so its properties under scale transformations are not immediately apparent.

In order to counter this, and perhaps to remove some of the confusion concerning the original argument of Cardy [1], it may be worth reiterating it in slightly different language. There are two important aspects to this argument. One is the assumption of scale invariance *in the continuum limit*, that is for properties which become independent of the lattice spacing as it is taken to zero. This is a well-accepted property of critical systems which should also hold for SAWs either in the fixed fugacity ensemble at the critical value of the fugacity, or deep inside large but finite SAWs of fixed length $N$. The other, however, is a careful identification of the continuum limit of the operator which counts the number of parallel contacts. It is this aspect of the argument which appears not to have been appreciated by the authors of Ref. [10]. Let us begin on the lattice, therefore, by defining a quantity $J_\mu^{\rm lat}(R)$ ($\mu = x$ or $y$) on each bond $R$, which takes the values 0, $+1$ or $-1$ according to whether the bond is unoccupied, or occupied by a bond parallel or antiparallel to the corresponding axis, respectively. $J_\mu^{\rm lat}$ is clearly a conserved current on the lattice, and its continuum version $J_\mu$ must also be conserved. This is defined so that the integral of the current $J_\mu$ flowing across a line segment is equal to the sum of the lattice currents flowing across the same segment. Thus $J_\mu^{\rm lat}$ and $J_\mu$ are related by a factor of the lattice spacing $b$, and $J_\mu$ naively has dimension $(\text{length})^{-1}$. It is an important property of conserved currents in continuum field theory that they retain their naive scaling dimension. This is because when integrated they must always yield a dimensionless charge (which in this case is the number of upgoing minus the number of downgoing arrows.)

Now the lattice operator $J_\mu^{\rm lat}(R)J_\mu^{\rm lat}(R')$, where $(R,R')$ are bonds on opposite sides of a lattice square, clearly counts the local density of parallel, minus the density of antiparallel contacts. The difference in these numbers for a given SAW is then given by summing over all such pairs $(R,R')$. In order to understand how this behaves for large SAWs we must therefore study the continuum limit of this operator. In general this may be a linear combination of all the possible local operators in the continuum theory (in fact, only those which, in the $O(n)$ mapping of the model, are symmetric under $O(n)$ rotations may appear). In order of increasing scaling dimensions, these are the unit operator, which cannot appear since it is nonzero even when there is no SAW in the vicinity, the energy density, and $:J^2:$. The energy density couples to the fugacity for each step in the walk, and it counts the number of local contacts, irrespective of their orientation. If such a term were added to the Hamiltonian, the critical value of the fugacity would change, corresponding to the fact that the mean number of such contacts grows like $N$. Since, for our model, we know this is not the case when we add an energy proportional to the number of parallel contacts only, we conclude that the energy density cannot enter into the continuum limit of the operator which counts parallel contacts only. This leaves $:J^2:$. This is the operator generated by the operator product expansion of two continuum currents

$$J(r)J(0) \sim \frac{\text{const.}}{r^2} + :J^2: + \cdots \tag{2}$$

where vector indices have been suppressed. It is a particular property of *two-dimensional* continuum field theory that $:J^2:$ retains its naive scaling dimension of $(\text{length})^{-2}$.

What this means is that, in the critical fugacity ensemble, or deep inside a long but finite SAW, the total number of parallel contacts inside a region $\mathcal{R}$, containing many lattice sites, is proportional to $\int_{\mathcal{R}} :J^2: d^2r$, and is actually scale invariant. Thus the number of parallel contacts within this region is $O(1)$, *independent* of the size of the region, as long as we stay away from the ends of the walk! This is to be contrasted with the argument of Ref. [10] quoted above, which would suggest that the number would depend strongly both on the size of the region and the distance from one of the ends. For the ensemble of SAWs with ends fixed at $r_1$, $r_2$ the distribution of parallel contacts is given, in the continuum limit, by the ratio of the correlators $\langle S(r_1) : J^2(r) : S(r_2)\rangle$ and $\langle S(r_1)S(r_2)\rangle$, where $S$ is the magnetisation operator of the $O(n)$ model. In the critical fugacity ensemble this is fixed by conformal invariance to have the form

$$\frac{\text{const.}|r_1 - r_2|^2}{|r - r_1|^2|r - r_2|^2} \tag{3}$$

We see that this is indeed approximately constant far from the ends, but that its integral exhibits a logarithmic divergence as they are approached, so that the total number of parallel contacts behaves as $\ln(|r_1 - r_2|/b)$, translating into a $\ln N$ dependence in the fixed $N$ ensemble.

We have repeated these arguments in detail in order to show that the derivation of the scaling behaviour of the number of parallel contacts requires several deep results in continuum field theory and is not simply given by naive scaling arguments. While these field theoretic arguments are by no means rigorous, they are generally accepted. Indeed the result that critical exponents should vary when a term proportional to $:J^2:$ is added to the Hamiltonian is directly responsible for the well-known non-universality exhibited by the two-dimensional XY model and the 1+1-dimensional Luttinger model.

It is therefore necessary to attempt to explain why the Monte Carlo simulations have apparently found no evidence for the logarithmic growth of the number of parallel contacts. One possible scenario is that the length of the walks so far considered $O(10^3)$ is still too short to observe the asymptotic behaviour. An argument for this may be advanced, based on the asymptotic relation between parallel contacts and the winding angle. This is perhaps most easily understood by considering the conformal transformation $z \to w = (L/2\pi)\ln z$ to the cylinder on which the coordinates are $L\ln r$ and $L\theta/2\pi$. An SAW winding around the origin corresponds to one winding around the cylinder. Scale invariance implies translational invariance along the cylinder, and that the typical number of windings in a segment of length $O(L)$ is $\pm O(1)$. The self-avoiding property of the walks also implies that the winding angles in two segments far apart on the cylinder are statistically uncorrelated, so that, by the central limit theorem, the total winding angle for a portion of the cylinder of length $L\ln r$ has a normal distribution with a width $O((\ln r)^{1/2})$ for large $\ln r$. This is the well-known result for the distribution of winding angles $\theta$. The width has been calculated exactly by Coulomb gas methods [20], giving $\langle\theta^2\rangle \sim 2\ln(N/b)$, where $b \approx 3.37$ is found from fitting to smaller values of $N$.

Turning to the occurrence of parallel contacts in the cylindrical geometry, the above field theory argument suggests that in a segment of length $O(L)$, there should be $O(1)$ such contacts. The same statistical argument shows that a segment of length $O(L\ln r)$ will then have $O(\ln r)$ contacts, for $\ln r \gg 1$. This is of course completely consistent with our result. However, it is clear that walks which do not backtrack along the cylinder must have a winding angle of $O(\pm 2\pi)$ if they are to have at least one parallel contact. Of course, once they have made such a contact, it is relatively easy for them to make further ones within a finite number of lattice spacings. However, in the continuum limit, all of these will be indistinguishable, their number simply renormalising the unknown coefficient multiplying the continuum operator $:J^2:$. It is not, therefore, these type of repeated parallel contacts which are responsible for the predicted $\ln N$ growth law: rather it is those which may occur each time the walk increases its winding angle by $\pm 2\pi$. If we now assume that a fraction $O(1)$ of such walks will have $O(1)$ new parallel contacts of this type we see that there must be at least $N \sim 3.37e^{2\pi^2} \approx 10^9$ steps in the walk in order to achieve just one new parallel contact. In order to observe a linear growth law with $\ln N$ one would need several powers of at least $10^9$. This is clearly outside the reach of current simulation techniques! The current walks of length $10^3$ have not (typically) rotated even by $\pi$ so will have very few renormalised parallel contacts. It would be interesting to test this scenario by measuring the correlations between parallel contacts and winding angle, or alternatively, to select from the ensemble only those walks which wind by at least $\pm 2\pi$. There would be greater statistical errors in such a procedure, however.

## Acknowledgements

We are grateful to Andrea Pelissetto for discussing his Monte Carlo results with us prior to their publication. Financial support from the Australian Research Council is gratefully acknowledged, and the Engineering and Physical Sciences Research Council under grant GR/J78327. One of us, D. B-W., thanks the Graduate School of the University of Melbourne for a scholarship.

## A First Correction Term for the Manhattan Lattice

The maximum perimeter of rectangles considered using the finite lattice method on the Manhattan lattice [12] up to width $W_{\max}$ is $4W_{\max}+4$ (It considers all such rectangles). It is then clear that the method up to this width can correctly find all the polygons up to length $4W_{\max}+4$. It will also include all polygons of length $4W_{\max}+8$ that can fit into rectangles of perimeter $4W_{\max}+4$ (and smaller). Clearly, the only polygons of length $4W_{\max}+8$ on the Manhattan lattice that cannot fit into these rectangles are the convex polygons of that length (There are no polygons that fit into rectangles of perimeter $4W_{\max}+6$, while not contained in ones of perimeter $4W_{\max}+4$, on the Manhattan lattice).

Convex polygons of length $4W_{\max}+8$ on the Manhattan lattice can be put into bijection with convex polygons of length $2W_{\max}+6$ on the square lattice in the following way. Consider any convex polygon of length $4W_{\max}+8$ on the Manhattan lattice with its bounding rectangle. Any straight section of the the polygon will contain an even number of steps except the 4 pieces that lie on the bounding rectangle which must have an odd number of steps (see figure 1). These constraints occur due to the Manhattan lattice orientations. Hence one can consider a map $\mathcal{M}$ that halves the length of each straight section of such a polygon except the parts lying in the boundary. The map takes these pieces lying in the boundary of length $2\ell+1$ and returns a section of length $\ell+1$. From these pieces a new convex polygon can be constructed and it will be of length $2W_{\max}+6$. Importantly the self-avoidance constraint is automatically satisfied when reconstructing since the polygon obtained is convex (see figure 2). The inverse map $\mathcal{M}^{-1}$ certainly exists and so this correspondence is one-to-one and onto. Note that an attempt to provide a similar mapping for 'almost convex' polygons on the Manhattan lattice, in principle which would allow further correction terms, is complicated by the self-avoidance constraint.

## B Field Theory Argument for Manhattan and L Lattices

It is possible to interpolate continuously between the ordinary square and the Manhattan or L lattice by adding a term to the Hamiltonian of the form

$$\lambda \sum_{i,j} \left( J_x^{\text{lat}}(i,j)(-1)^j + J_y^{\text{lat}}(i,j)(-1)^i \right) \tag{4}$$

for the Manhattan lattice, and

$$\lambda \sum_{i,j} \left( J_x^{\rm lat}(i,j) + J_y^{\rm lat}(i,j) \right) (-1)^{i+j} \tag{5}$$

for the L lattice. In each case, $\lambda = 0$ gives the usual square lattice and $\lambda \to \infty$ corresponds to the fully directed lattice.

To first order in $\lambda$ the effect of these oscillating terms averages to zero in any physical quantity. However, to second order there are unmodulated terms which will survive in the continuum limit, when they will again be proportional to the operator $:J^2:$ and hence give rise to a continuously varying $\gamma$. For small $\lambda$ this effect will be even weaker than that predicted for the square lattice with parallel interactions, since it is $O(\lambda^2)$, but for infinite $\lambda$ it is expected to be of the same order of magnitude.

## References

[1] J. L. Cardy, Nuc. Phys. B. **419**, 411 (1994).

[2] D. Bennett-Wood, J. L. Cardy, S. Flesia, A. J. Guttmann, and A. L. Owczarek, J. Phys. A. **28**, 5143 (1995).

[3] A. Malakis, J. Phys. A. **8**, 1885 (1975).

[4] P. Grassberger, Z. Phys. B **48**, 255 (1982).

[5] A. J. Guttmann, J. Phys. A. **16**, 3894 (1983).

[6] S. Flesia, Europhys. Lett. **32**, 149 (1995).

[7] W. M. Koo, J. Stat. Phys. **81**, 561 (1995).

[8] G. T. Barkema and S. Flesia, J. Stat. Phys. **85**, 363 (1996).

[9] A. Trovato and F. Seno, Phys. Rev. E **56**, 131 (1997).

[10] T. Prellberg and B. Drossel, Preprint, 1997.

[11] G. T. Barkema, U. Bastolla, and P. Grassberger, preprint, 1997.

[12] I. G. Enting and A. J. Guttmann, J. Phys. A **18**, 1007 (1985).

[13] B. Nienhuis, Phys. Rev. Lett. **49**, 1062 (1982).

[14] I. G. Enting, J. Phys. A **13**, 3713 (1980).

[15] A. J. Guttmann, in *Phase Transitions and Critical Phenomena*, edited by C. Domb and J. L. Lebowitz, volume 13, Academic Press, 1989.

[16] A. Conway and A. J. Guttmann, Phys. Rev. Lett. **77**, 5284 (1996).

[17] A. Conway, I. G. Enting, and A. J. Guttmann, J. Phys. A. **26**, 1519 (1993).

[18] A. Pelissetto, Private communication.

[19] B. Li, N. Madras, and A. D. Sokal, J. Stat. Phys. **80**, 661 (1995).

[20] B. Duplantier and F. David, J. Stat. Phys. **51**, 327 (1988).

## Table Captions

Table 1.

The number of walks $w_n$ on the Manhattan lattice up to length $n = 53$.

Table 2.

The number of polygons $p_n$ on the Manhattan lattice up to length $n = 84$.

Table 3.

A fit to the even subsequence with $\gamma = 1.3385$, and $\Theta = 1.54$ and $\Delta = 1.5$ and $\Lambda = 1$.

Table 4.

A fit to the odd subsequence with $\gamma = 1.3385$, and $\Theta = 1.54$ and $\Delta = 1.5$ and $\Lambda = 1$.

Table 5.

A fit to the even subsequence with $\gamma = 1.34375$, and $\Theta = 1.54$ and $\Delta = 1.5$ and $\Lambda = 1$.

Table 6.

A fit to the even subsequence with $x_c = 0.576857, \gamma = 1.3385$, and $\Theta = 1.54$ and $\Delta = 1.5$ and $\Lambda = 1$.

Table 7.

A fit to the even subsequence with $\gamma = 1.3385$, and $\Theta = 1.50$ and $\Delta = 1.5$ and $\Lambda = 1$.

Table 8.

A fit to the asymptotic form with $\gamma = 43/32$.

Table 9.

A fit to the asymptotic form with $\gamma = 1.3394$ and an antiferromagnetic exponent 0.50.

Table 10.

A fit of the two-legged star data to the asymptotic form with $\gamma^{(2)} = 2.3215$ and an antiferromagnetic exponent 1.50.

## Figure Captions

Figure 1.

The two different types of 'straight section' in a polygon on the Manhattan lattice: type 1 is always of odd length and type 2 is always of even length. A convex polygon only has four type-1 sections but may have any number of type-2 sections.

Figure 2.

A convex polygon on the Manhattan lattice and its counterpart on the square lattice linked by the map $\mathcal{M}$ described in the text of the Appendix. On either lattice the four directed walks connecting the type-1 sections are self-avoiding by construction. Hence, the map $\mathcal{M}$ from one convex structure to another always obeys self-avoidance.

Figure 3.

A non-backtracking walk on the cylinder with a parallel contact must have a winding angle of $\pm 2\pi$ in order to achieve this. Once this happens, further parallel contacts are more likely to occur within a finite number of lattice spacings.

| $n$ | $w_n/2$ | $n$ | $w_n/2$ | $n$ | $w_n/2$ |
|---|---|---|---|---|---|
| 1 | 1 | 19 | 43106 | 37 | 1058544744 |
| 2 | 2 | 20 | 75396 | 38 | 1852200487 |
| 3 | 4 | 21 | 132865 | 39 | 3241111183 |
| 4 | 7 | 22 | 234171 | 40 | 5653133990 |
| 5 | 13 | 23 | 412731 | 41 | 9881311436 |
| 6 | 24 | 24 | 721433 | 42 | 17273983512 |
| 7 | 44 | 25 | 1267901 | 43 | 30199278540 |
| 8 | 77 | 26 | 2228666 | 44 | 52652493201 |
| 9 | 139 | 27 | 3917654 | 45 | 91964600384 |
| 10 | 250 | 28 | 6843596 | 46 | 160645950194 |
| 11 | 450 | 29 | 12004150 | 47 | 280636185403 |
| 12 | 788 | 30 | 21059478 | 48 | 489116742528 |
| 13 | 1403 | 31 | 36947904 | 49 | 853776966616 |
| 14 | 2498 | 32 | 64506130 | 50 | 1490455491081 |
| 15 | 4447 | 33 | 112983428 | 51 | 2602057537031 |
| 16 | 7782 | 34 | 197921386 | 52 | 4533660722293 |
| 17 | 13769 | 35 | 346735329 | 53 | 7909561970564 |
| 18 | 24363 | 36 | 605046571 | | |

Table 1: The number of walks $w_n$ on the Manhattan lattice up to length $n = 53$.

| $n$ | $p_n/2$ |
|---|---|
| 4 | 1 |
| 8 | 2 |
| 12 | 7 |
| 16 | 32 |
| 20 | 168 |
| 24 | 970 |
| 28 | 5984 |
| 32 | 38786 |
| 36 | 261160 |
| 40 | 1812630 |
| 44 | 12895360 |
| 48 | 93638634 |
| 52 | 691793872 |
| 56 | 5186869122 |
| 60 | 39388514522 |
| 64 | 302457399674 |
| 68 | 2345362579172 |
| 72 | 18345337742960 |
| 76 | 144612959167806 |
| 80 | 1147920496989270 |
| 84 | 9169516892088470 |

Table 2: The number of polygons $p_n$ on the Manhattan lattice up to length $n = 84$.

| $n$ | $a_1^e$ | $a_2^e$ | $a_3^e$ | $a_4^e$ | $f_1^e$ | $f_2^e$ | $f_3^e$ |
|---|---|---|---|---|---|---|---|
| 14 | 1.11532 | 0.63062 | -1.11528 | 0.70101 | -0.16447 | 0.18397 | 0.3570 |
| 15 | 1.11381 | 0.74606 | -1.76108 | 1.41788 | -0.15574 | -0.01487 | 1.4778 |
| 16 | 1.11405 | 0.72571 | -1.64241 | 1.28234 | -0.15421 | -0.05281 | 1.7110 |
| 17 | 1.11404 | 0.72648 | -1.64709 | 1.28782 | -0.15415 | -0.05436 | 1.7213 |
| 18 | 1.11410 | 0.72095 | -1.61229 | 1.24601 | -0.15374 | -0.06623 | 1.8063 |
| 19 | 1.11397 | 0.73406 | -1.69747 | 1.35071 | -0.15277 | -0.09611 | 2.0355 |
| 20 | 1.11391 | 0.74055 | -1.74098 | 1.40533 | -0.15324 | -0.08044 | 1.9074 |
| 21 | 1.11389 | 0.74336 | -1.76038 | 1.43018 | -0.15304 | -0.08760 | 1.9695 |
| 22 | 1.11389 | 0.74268 | -1.75558 | 1.42391 | -0.15299 | -0.08942 | 1.9862 |
| 23 | 1.11389 | 0.74369 | -1.76297 | 1.43374 | -0.15292 | -0.09227 | 2.0138 |
| 24 | 1.11388 | 0.74454 | -1.76926 | 1.44224 | -0.15298 | -0.08979 | 1.9886 |
| 25 | 1.11390 | 0.74147 | -1.74576 | 1.40996 | -0.15320 | -0.08037 | 1.8878 |
| 26 | 1.11391 | 0.74035 | -1.73702 | 1.39777 | -0.15312 | -0.08393 | 1.9277 |

Table 3: A fit to the even subsequence with $\gamma = 1.3385$, and $\Theta = 1.54$ and $\Delta = 1.5$ and $\Lambda = 1$.

| $n$ | $a_1^o$ | $a_2^o$ | $a_3^o$ | $a_4^o$ | $f_1^o$ | $f_2^o$ | $f_3^o$ |
|---|---|---|---|---|---|---|---|
| 14 | 1.93367 | 1.39292 | -1.71641 | 1.08024 | -0.47972 | 0.80137 | 0.2368 |
| 15 | 1.93256 | 1.47690 | -2.18623 | 1.60176 | -0.47337 | 0.65671 | 1.0522 |
| 16 | 1.93217 | 1.50941 | -2.37590 | 1.81839 | -0.47581 | 0.71735 | 0.6795 |
| 17 | 1.93199 | 1.52502 | -2.47052 | 1.92934 | -0.47465 | 0.68606 | 0.8877 |
| 18 | 1.93202 | 1.52238 | -2.45395 | 1.90942 | -0.47445 | 0.68041 | 0.9282 |
| 19 | 1.93182 | 1.54283 | -2.58686 | 2.07278 | -0.47294 | 0.63378 | 1.2857 |
| 20 | 1.93173 | 1.55224 | -2.64993 | 2.15197 | -0.47363 | 0.65649 | 1.1001 |
| 21 | 1.93171 | 1.55466 | -2.66662 | 2.17334 | -0.47345 | 0.65033 | 1.1535 |
| 22 | 1.93167 | 1.56012 | -2.70537 | 2.22393 | -0.47385 | 0.66497 | 1.0192 |
| 23 | 1.93170 | 1.55575 | -2.67352 | 2.18160 | -0.47417 | 0.67725 | 0.9002 |
| 24 | 1.93168 | 1.55782 | -2.68904 | 2.20258 | -0.47432 | 0.68336 | 0.8380 |
| 25 | 1.93176 | 1.54783 | -2.61255 | 2.09750 | -0.47503 | 0.71404 | 0.5100 |
| 26 | 1.93177 | 1.54609 | -2.59894 | 2.07852 | -0.47491 | 0.70849 | 0.5722 |

Table 4: A fit to the odd subsequence with $\gamma = 1.3385$, and $\Theta = 1.54$ and $\Delta = 1.5$ and $\Lambda = 1$.

| $n$ | $a_1^e$ | $a_2^e$ | $a_3^e$ | $a_4^e$ | $f_1^e$ | $f_2^e$ | $f_3^e$ |
|---|---|---|---|---|---|---|---|
| 14 | 1.08859 | 1.03022 | -2.54569 | 1.97617 | -0.16591 | 0.21526 | 0.1889 |
| 15 | 1.08660 | 1.18063 | -3.37952 | 2.90520 | -0.15431 | -0.04877 | 1.6772 |
| 16 | 1.08639 | 1.19776 | -3.47856 | 3.01878 | -0.15563 | -0.01619 | 1.4770 |
| 17 | 1.08597 | 1.23562 | -3.70602 | 3.28671 | -0.15274 | -0.09363 | 1.9923 |
| 18 | 1.08564 | 1.26731 | -3.90332 | 3.52493 | -0.15514 | -0.02433 | 1.4961 |
| 19 | 1.08515 | 1.31726 | -4.22495 | 3.92232 | -0.15137 | -0.14057 | 2.3873 |
| 20 | 1.08474 | 1.36071 | -4.51365 | 4.28684 | -0.15463 | -0.03343 | 1.5117 |
| 21 | 1.08440 | 1.40056 | -4.78632 | 4.63816 | -0.15166 | -0.13714 | 2.4116 |
| 22 | 1.08410 | 1.43700 | -5.04267 | 4.97487 | -0.15436 | -0.03736 | 1.4955 |
| 23 | 1.08380 | 1.47509 | -5.31782 | 5.34289 | -0.15155 | -0.14681 | 2.5554 |
| 24 | 1.08352 | 1.51302 | -5.59869 | 5.72510 | -0.15434 | -0.03278 | 1.3937 |
| 25 | 1.08328 | 1.54711 | -5.85731 | 6.08287 | -0.15184 | -0.13983 | 2.5382 |
| 26 | 1.08303 | 1.58312 | -6.13673 | 6.47552 | -0.15447 | -0.02203 | 1.2196 |

Table 5: A fit to the even subsequence with $\gamma = 1.34375$, and $\Theta = 1.54$ and $\Delta = 1.5$ and $\Lambda = 1$.

| $n$ | $a_1^e$ | $a_2^e$ | $a_3^e$ | $a_4^e$ | $f_1^e$ | $f_2^e$ | $f_3^e$ |
|---|---|---|---|---|---|---|---|
| 14 | 1.11554 | 0.62302 | -1.08221 | 0.66929 | -0.16443 | 0.18302 | 0.3624 |
| 15 | 1.11404 | 0.73699 | -1.71984 | 1.37709 | -0.15581 | -0.01330 | 1.4691 |
| 16 | 1.11430 | 0.71510 | -1.59212 | 1.23122 | -0.15416 | -0.05413 | 1.7200 |
| 17 | 1.11431 | 0.71416 | -1.58640 | 1.22451 | -0.15424 | -0.05224 | 1.7074 |
| 18 | 1.11439 | 0.70682 | -1.54028 | 1.16910 | -0.15369 | -0.06796 | 1.8201 |
| 19 | 1.11428 | 0.71796 | -1.61269 | 1.25810 | -0.15287 | -0.09336 | 2.0148 |
| 20 | 1.11424 | 0.72240 | -1.64245 | 1.29546 | -0.15319 | -0.08265 | 1.9272 |
| 21 | 1.11423 | 0.72299 | -1.64654 | 1.30069 | -0.15315 | -0.08416 | 1.9403 |
| 22 | 1.11426 | 0.72001 | -1.62539 | 1.27309 | -0.15293 | -0.09214 | 2.0136 |
| 23 | 1.11427 | 0.71856 | -1.61479 | 1.25900 | -0.15304 | -0.08805 | 1.9740 |
| 24 | 1.11428 | 0.71685 | -1.60200 | 1.24172 | -0.15291 | -0.09308 | 2.0252 |
| 25 | 1.11432 | 0.71106 | -1.55767 | 1.18083 | -0.15333 | -0.07530 | 1.8352 |
| 26 | 1.11435 | 0.70714 | -1.52699 | 1.13801 | -0.15305 | -0.08783 | 1.9755 |

Table 6: A fit to the even subsequence with $x_c = 0.576857, \gamma = 1.3385$, and $\Theta = 1.54$ and $\Delta = 1.5$ and $\Lambda = 1$.

| $n$ | $a_1^e$ | $a_2^e$ | $a_3^e$ | $a_4^e$ | $f_1^e$ | $f_2^e$ | $f_3^e$ |
|---|---|---|---|---|---|---|---|
| 14 | 1.11538 | 0.62667 | -1.09363 | 0.67732 | -0.14091 | 0.03836 | 0.7105 |
| 15 | 1.11377 | 0.74910 | -1.77850 | 1.43750 | -0.13250 | -0.15318 | 1.7905 |
| 16 | 1.11408 | 0.72295 | -1.62597 | 1.26331 | -0.13072 | -0.19731 | 2.0619 |
| 17 | 1.11401 | 0.72909 | -1.66316 | 1.30691 | -0.13031 | -0.20840 | 2.1357 |
| 18 | 1.11413 | 0.71851 | -1.59668 | 1.22704 | -0.12960 | -0.22880 | 2.2818 |
| 19 | 1.11395 | 0.73633 | -1.71246 | 1.36933 | -0.12842 | -0.26523 | 2.5612 |
| 20 | 1.11393 | 0.73836 | -1.72609 | 1.38644 | -0.12855 | -0.26084 | 2.5253 |
| 21 | 1.11387 | 0.74543 | -1.77494 | 1.44902 | -0.12809 | -0.27693 | 2.6649 |
| 22 | 1.11391 | 0.74069 | -1.74127 | 1.40506 | -0.12778 | -0.28825 | 2.7688 |
| 23 | 1.11387 | 0.74560 | -1.77703 | 1.45258 | -0.12747 | -0.30050 | 2.8875 |
| 24 | 1.11389 | 0.74270 | -1.75540 | 1.42335 | -0.12728 | -0.30804 | 2.9643 |
| 25 | 1.11389 | 0.74324 | -1.75947 | 1.42894 | -0.12725 | -0.30948 | 2.9798 |
| 26 | 1.11392 | 0.73865 | -1.72354 | 1.37881 | -0.12696 | -0.32245 | 3.1249 |

Table 7: A fit to the even subsequence with $\gamma = 1.3385$, and $\Theta = 1.50$ and $\Delta = 1.5$ and $\Lambda = 1$.

| $n$ | $a_1$ | $a_2$ | $a_3$ | $b_1$ | $b_2$ |
|---|---|---|---|---|---|
| 12 | 1.14916623 | 0.97877877 | -0.81158039 | -0.36375929 | 0.92859850 |
| 13 | 1.14143368 | 1.22320520 | -1.33919183 | -0.32200448 | 0.49321719 |
| 14 | 1.14509983 | 1.09635554 | -1.05260073 | -0.30059269 | 0.24840440 |
| 15 | 1.14267035 | 1.18768479 | -1.26774563 | -0.28534984 | 0.05879950 |
| 16 | 1.14224305 | 1.20502692 | -1.31020454 | -0.28821352 | 0.09729745 |
| 17 | 1.14204805 | 1.21352460 | -1.33176762 | -0.28682439 | 0.07722804 |
| 18 | 1.14111738 | 1.25686961 | -1.44549198 | -0.29384278 | 0.18566830 |
| 19 | 1.14091474 | 1.26691473 | -1.47268541 | -0.29223093 | 0.15914714 |
| 20 | 1.14050244 | 1.28858752 | -1.53311033 | -0.29567880 | 0.21933514 |
| 21 | 1.14008082 | 1.31201320 | -1.60026457 | -0.29198241 | 0.15110381 |
| 22 | 1.13980150 | 1.32836978 | -1.64840570 | -0.29454331 | 0.20094150 |
| 23 | 1.13948412 | 1.34790622 | -1.70736309 | -0.29150725 | 0.13881468 |
| 24 | 1.13918560 | 1.36717694 | -1.76692129 | -0.29448076 | 0.20264014 |
| 25 | 1.13894695 | 1.38329792 | -1.81789123 | -0.29201018 | 0.14713562 |
| 26 | 1.13868471 | 1.40179841 | -1.87767115 | -0.29482690 | 0.21323756 |
| 27 | 1.13847187 | 1.41745252 | -1.92931932 | -0.29245851 | 0.15528540 |
| 28 | 1.13824919 | 1.43449751 | -1.98669333 | -0.29502175 | 0.22057157 |
| 29 | 1.13806006 | 1.44954118 | -2.03831472 | -0.29277265 | 0.16103498 |
| 30 | 1.13786403 | 1.46572219 | -2.09487723 | -0.29517818 | 0.22712062 |
| 31 | 1.13769766 | 1.47995338 | -2.14552077 | -0.29307402 | 0.16720787 |
| 32 | 1.13752301 | 1.49541683 | -2.20150697 | -0.29534837 | 0.23424272 |
| 33 | 1.13737469 | 1.50899312 | -2.25148713 | -0.29336173 | 0.17369987 |
| 34 | 1.13721868 | 1.52374228 | -2.30666839 | -0.29550935 | 0.24129808 |

Table 8: A fit to the asymptotic form with $\gamma = 43/32$.

| $n$ | $a_1$ | $a_2$ | $a_3$ | $b_1$ | $b_2$ |
|---|---|---|---|---|---|
| 12 | 1.16934961 | 0.83811400 | -0.61621261 | -0.36250828 | 0.91618097 |
| 13 | 1.16199936 | 1.07045150 | -1.11772419 | -0.32321982 | 0.50650408 |
| 14 | 1.16612567 | 0.92768248 | -0.79516921 | -0.29937421 | 0.23385735 |
| 15 | 1.16405934 | 1.00535893 | -0.97815085 | -0.28655106 | 0.07434748 |
| 16 | 1.16398820 | 1.00824601 | -0.98521927 | -0.28702245 | 0.08068475 |
| 17 | 1.16412793 | 1.00215704 | -0.96976842 | -0.28800635 | 0.09489981 |
| 18 | 1.16350184 | 1.03131595 | -1.04627236 | -0.29267192 | 0.16698803 |
| 19 | 1.16359366 | 1.02676485 | -1.03395197 | -0.29339336 | 0.17885871 |
| 20 | 1.16345663 | 1.03396792 | -1.05403447 | -0.29452513 | 0.19861572 |
| 21 | 1.16329519 | 1.04293753 | -1.07974750 | -0.29312762 | 0.17281887 |
| 22 | 1.16326445 | 1.04473754 | -1.08504533 | -0.29340583 | 0.17823317 |
| 23 | 1.16318304 | 1.04974902 | -1.10016902 | -0.29263717 | 0.16250404 |
| 24 | 1.16310969 | 1.05448367 | -1.11480193 | -0.29335806 | 0.17797789 |
| 25 | 1.16308698 | 1.05601804 | -1.11965314 | -0.29312608 | 0.17276603 |
| 26 | 1.16303114 | 1.05995718 | -1.13238150 | -0.29371764 | 0.18664881 |
| 27 | 1.16301691 | 1.06100363 | -1.13583408 | -0.29356151 | 0.18282827 |
| 28 | 1.16298490 | 1.06345400 | -1.14408210 | -0.29392485 | 0.19208260 |
| 29 | 1.16297969 | 1.06386843 | -1.14550420 | -0.29386376 | 0.19046564 |
| 30 | 1.16296076 | 1.06543081 | -1.15096566 | -0.29409271 | 0.19675539 |
| 31 | 1.16296569 | 1.06500942 | -1.14946610 | -0.29415411 | 0.19850378 |
| 32 | 1.16295638 | 1.06583321 | -1.15244867 | -0.29427351 | 0.20202281 |
| 33 | 1.16296836 | 1.06473676 | -1.14841215 | -0.29443158 | 0.20684031 |
| 34 | 1.16296742 | 1.06482601 | -1.14874608 | -0.29444439 | 0.20724330 |

Table 9: A fit to the asymptotic form with $\gamma = 1.3394$ and an antiferromagnetic exponent 0.50.

| $n$ | $a_1$ | $a_2$ | $a_3$ | $b_1$ | $b_2$ |
|---|---|---|---|---|---|
| 12 | 0.22309468 | 0.59133069 | -0.09371775 | -0.03242810 | 0.07726904 |
| 13 | 0.22135318 | 0.64637384 | -0.21252607 | -0.02350139 | -0.01582495 |
| 14 | 0.22178976 | 0.63126937 | -0.17840204 | -0.02108596 | -0.04344544 |
| 15 | 0.22115378 | 0.65517534 | -0.23471546 | -0.01731317 | -0.09038006 |
| 16 | 0.22100884 | 0.66105764 | -0.24911669 | -0.01823004 | -0.07805290 |
| 17 | 0.22087410 | 0.66692907 | -0.26401520 | -0.01732551 | -0.09112218 |
| 18 | 0.22069229 | 0.67539647 | -0.28623061 | -0.01861564 | -0.07118719 |
| 19 | 0.22060737 | 0.67960578 | -0.29762553 | -0.01798096 | -0.08163072 |
| 20 | 0.22055168 | 0.68253291 | -0.30578637 | -0.01841795 | -0.07400191 |
| 21 | 0.22046345 | 0.68743505 | -0.31983906 | -0.01769298 | -0.08738487 |
| 22 | 0.22044332 | 0.68861361 | -0.32330779 | -0.01786572 | -0.08402302 |
| 23 | 0.22040353 | 0.69106274 | -0.33069871 | -0.01750981 | -0.09130610 |
| 24 | 0.22038824 | 0.69204993 | -0.33374970 | -0.01765210 | -0.08825185 |
| 25 | 0.22037687 | 0.69281811 | -0.33617842 | -0.01754225 | -0.09071992 |
| 26 | 0.22037678 | 0.69282412 | -0.33619785 | -0.01754310 | -0.09069989 |
| 27 | 0.22037843 | 0.69270277 | -0.33579748 | -0.01756020 | -0.09028147 |

Table 10: A fit of the two-legged star data to the asymptotic form with $\gamma^{(2)} = 2.3215$ and an antiferromagnetic exponent 1.5.

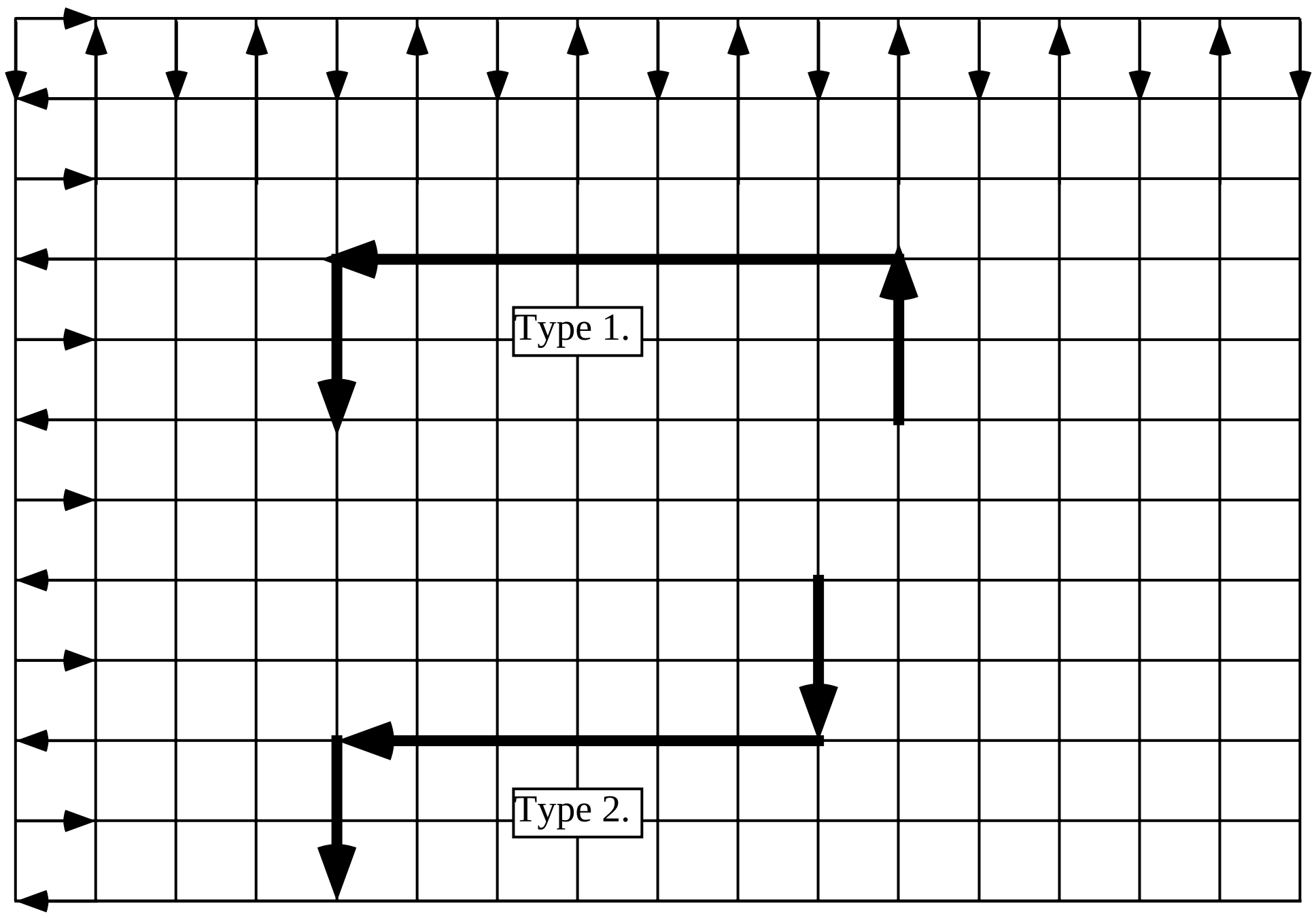

Figure 1:

**Title:** On the Non-Universality of a Critical Exponent for Self-Avoiding Walks

**Authors:** D. Bennett-Wood *et. al.*

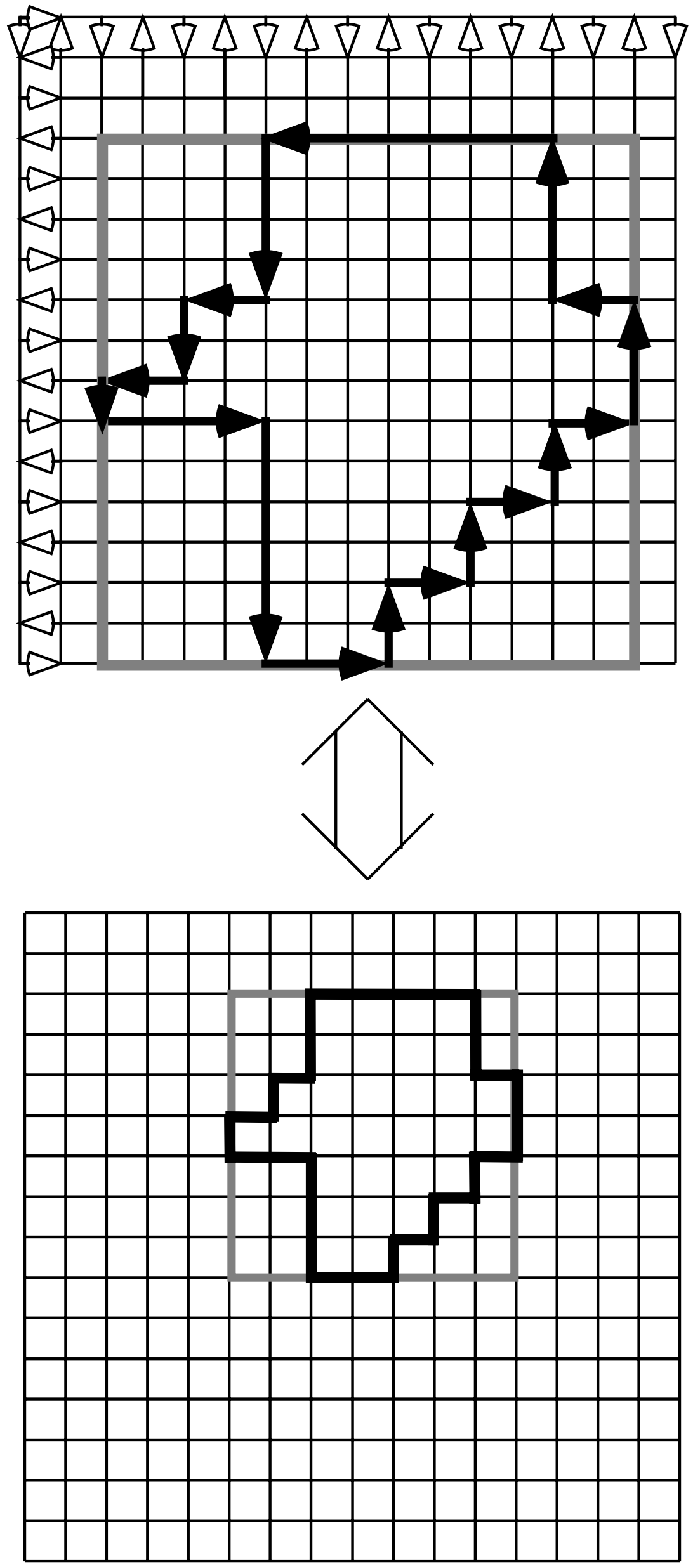

Figure 2:

**Title:** On the Non-Universality of a Critical Exponent for Self-Avoiding Walks

**Authors:** D. Bennett-Wood *et. al.*

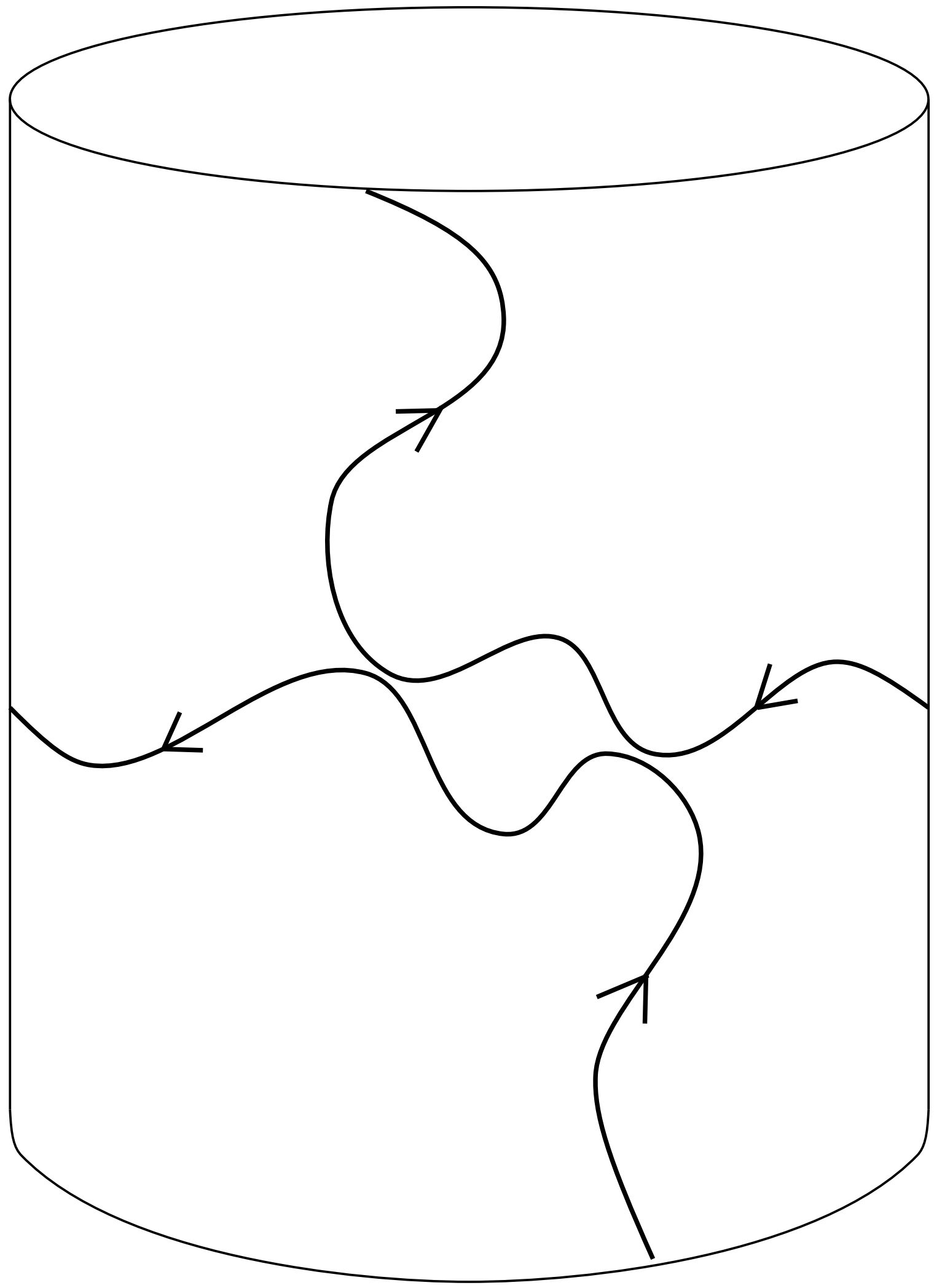

Figure 3:

**Title:** On the Non-Universality of a Critical Exponent for Self-Avoiding Walks

**Authors:** D. Bennett-Wood *et. al.*